\documentclass[final,3p,times,authoryear]{elsarticle}

\usepackage{blkarray}

\usepackage{amssymb}
\usepackage{amsmath}
\usepackage{graphicx}
\usepackage{caption} 
\usepackage{subfig} 
\usepackage{float} 
\usepackage{hyperref}
\usepackage{algorithm}
\usepackage{algpseudocode}
\usepackage{multirow}

\graphicspath{{Results/}}

\label{sec.sub.gdn}

\newtheorem{Theorem}{Theorem}[section]
\newtheorem{Definition}[Theorem]{Definition}

\newtheorem{Corollary}[Theorem]{Corollary}
\newtheorem{Lemma}[Theorem]{Lemma}
\newtheorem{Example}[Theorem]{Example}
\newtheorem{Remark}[Theorem]{Remark}
\newenvironment{Proof}{{\bf \noindent Proof.}\rm}{\hfill{$\Box$}}

\journal{Group Decision and Negotiation}

\makeatletter
\newcommand*\bigcdot{\mathpalette\bigcdot@{.5}}
\newcommand*\bigcdot@[2]{\mathbin{\vcenter{\hbox{\scalebox{#2}{$\m@th#1\bullet$}}}}}
\makeatother

\date{}

\begin{document}

\begin{frontmatter}




\title{Unveiling and unraveling aggregation and dispersion fallacies in group MCDM}

\author[vu]{Majid Mohammadi\corref{cor1}}
\ead{majid.mohammadi690@gmail.com.}
\author[jads]{Damian A. Tamburri}
\author[delft]{Jafar Rezaei}
\cortext[cor1]{Corresponding author}
\address[vu]{Department of Computer Science, Vrije Universiteit Amsterdam, The Netherlands}
\address[jads]{Jheronimus Academy of Data Science, Eindhoven University of Technology, The Netherlands}
\address[delft]{Faculty of Technology, Policy and Management, Delft University of Technology, The Netherlands}

\begin{abstract}
Priorities in multi-criteria decision-making (MCDM) convey the relevance preference of one criterion over another, which is usually reflected by imposing the non-negativity and unit-sum constraints. The processing of such priorities is different than other unconstrained data, but this point is often neglected by researchers, which results in fallacious statistical analysis. This article studies three prevalent fallacies in group MCDM along with solutions based on compositional data analysis to avoid misusing statistical operations. First, we use a compositional approach to aggregate the priorities of a group of DMs and show that the outcome of the compositional analysis is identical to the normalized geometric mean, meaning that the arithmetic mean should be avoided. Furthermore, a new aggregation method is developed, which is a robust surrogate for the geometric mean. We also discuss the errors in computing measures of dispersion, including standard deviation and distance functions. Discussing the fallacies in computing the standard deviation, we provide a probabilistic criteria ranking by developing proper Bayesian tests, where we calculate the extent to which a criterion is more important than another. Finally, we explain the errors in computing the distance between priorities, and a clustering algorithm is specially tailored based on proper distance metrics.
\end{abstract}

\end{frontmatter}

\section{Introduction}
Multi-criteria decision-making (MCDM) problems typically involve evaluating a set of alternatives with respect to a handful of criteria based on the preferences of one or a group of decision-makers (DMs), with the ultimate goal of selecting, sorting, or ranking available alternatives. For such evaluation, the performance of alternatives for each criterion is acquired by employing a crucial data collection approach, whose results are stored in a so-called \textit{performance matrix}. There are several methods to elicit the preferences of DMs, including but not limited to Tradeoff \citep{Keeney}, SMART (simple multi-attribute rating technique) \citep{smart}, Swing \citep{swing}, AHP (analytic hierarchy process) \citep{ahp}, ANP (analytic network process) \citep{anp}, and BWM (best-worst method) \citep{bwm}. For more information about popular MCDM methods, see \citep{mcdm_book}.

Conventional MCDM methods typically analyze a decision-making problem that entails one DM only. Members usually have distinct preferences when there is a group of DMs for decision-making. Three approaches exist to deal with the differences in preferences: sharing, comparing, and aggregating \citep{belton1997}. In sharing, the whole group will arrive at a unified preference structure so that the group is treated as a single DM problem. For instance, if one wants to find the weights of a set of criteria in sharing, they will end up with a single set of weights through negotiation among members and by using a weighting method (e.g., AHP).
In contrast, individual preferences are considered in comparing and aggregating, where we have different sets of weights from individual members. In aggregating, for instance, we try to aggregate the weights from all members to come up with a single set of weights \citep{fuckingReferee}.

Aggregating is usually performed in two different ways. First, the methods for one DM are extended to encompass multiple DMs' preferences. The most popular technique known so far is arguably the geometric mean method {\citep{aggregate_1994,aggregate_1998}} that is applied, for instance, to the pairwise comparison matrices (PCMs) computed within the AHP, the result of which is an aggregated PCM representing the preferences of the whole group. Other methods use more complicated techniques, such as evolutionary algorithms \citep{aij_ea1,aij_ea2} and Bayesian statistics \citep{bayesian_ahp,bayesian_bwm}, to aggregate the preferences of multiple DMs. This is also called input-based aggregation \citep{dias2005dealing}. The second approach for aggregation is to compute all DMs' priorities individually and then conduct the consequent processes based on such priorities, such as aggregation and clustering of DMs. This approach is called output-based aggregation \citep{dias2005dealing}. The arithmetic mean is typically used for output-based aggregation and is seemingly appropriate since it satisfies the non-negativity and constant-sum (e.g., unit-sum) constraints required in most MCDM methods. Other processes, such as computing the standard deviation of priorities of the group \citep{fal_standardDeviation_1,fal_standardDeviation_2}, clustering of DMs \citep{fal_clustering_1,fal_clustering_2}, and statistical significance tests \citep{fal_statest_1,aij_ea1}, are directly applied to the priorities computed based on the preferences of DMs.

While the methods for aggregating multiple DMs' preferences are often statistically sound, applying the same statistical operations to the priorities is incorrect because the priorities are ratios lying on a simplex and not on the real space. Like the priorities of a DM, a vector that satisfies the non-negativity and constant-sum constraints is called a \textit{composition}\footnote{Even if they do not lie on a standard simplex, the priorities in MCDM convey the relative importance of one criterion over another, meaning that they are still compositional and statistical methodologies for processing them should be different than other standard multivariate data.}\citep{compositional_book}. From a statistical point of view, the statistical operations should be adjusted to apply to such compositions; otherwise, the follow-up methods or outcomes are unreliable and statistically incorrect. While there is a tremendous effort, such as several books \citep{compositional_book,compositional_book2,compositional_book3, compositional_ranking}, in providing proper statistical tools and methods for analyzing compositional data, MCDM researchers often neglect them, resulting in improper statistical analysis with disastrous consequences. This article aims to study three of such essential and prevalent fallacies in analyzing the priorities of a group of DMs.

First, we analyze the aggregation of different DMs' priorities using the notions of compositional data since computing the arithmetic or geometric mean directly from the priorities is not in line with their compositional nature. We show that the aggregated priorities based on our analysis are equivalent to the normalized geometric mean of the priorities, called the geometric mean method (GMM), in the MCDM literature \citep{aggregate_1998,aggregate_1994}. The byproduct of such an analysis is that using the arithmetic mean for aggregating the priorities should be avoided. This analysis ends a lengthy discussion in MCDM regarding using the arithmetic or geometric mean for aggregating priorities. In addition, the adaptive weighted geometric mean method (AWGMM) is developed, which is a more robust surrogate to the GMM. The robustness of the method is against DMs whose preferences significantly deviate from the majority of other DMs. The AWGMM uses the robust Welsch estimator developed in robust statistics \citep{robust_statistics}, which adaptively assigns a weight for each DM based on their priorities. A DM with a small weight is deemed \textit{deviant} and thus contributes less to the final aggregated priorities computed in the AWGMM. Aside from the aggregation, the identification of DMs with deviating preferences could also be used in the negotiation process to converge the preferences of the whole group.  

Second, the errors in estimating priorities' standard deviation (as a measure of dispersion) are explained. Since statistical tests are typically based on the standard deviation of samples, the flaw in computing the standard deviation directly impacts the tests. Thus, the change in calculating the standard deviation for priorities (as compositional data) means that statistical tests cannot be directly applied to priorities to verify the difference between the weights of the two criteria. For example, statistical tests, such as paired \textit{t}-test and Wilcoxon Signed-rank test, are used to check the importance difference of criteria based on the priorities of a group by \textit{subtracting} their weights \citep{fal_statest_1}, which is in not statistically correct given the compositional nature of the priorities. Instead, we develop a Bayesian Wilcoxon-type test to verify if the difference between the weights of two criteria is significant based on a group of DMs' priorities. The Bayesian tests do not suffer from the pitfalls of the frequentist tests and enable us to compute the extent to which one criterion is preferred over another based on the priorities of a group of DMs. This approach provides us with a probabilistic ranking of criteria. Aside from the Wilcoxon-type test, the ways to use the Bayesian \textit{t}-test and Bayesian Sign test (i.e., beta-binomial conjugate) are also explained.  

Third, we put forward the proper ways to measure the distance (as another measure of dispersion) between two priorities. One of the most popular distance functions is the Euclidean distance which is directly applied to the weights in the priorities \citep{fal_clustering_1}. The distance functions are used for different aims, but one of the essential uses of a distance function is clustering the DMs based on their priorities. Typically, the Euclidean distance and mean absolute deviation (MAD) are used for clustering the DMs using clustering algorithms such as K-means and fuzzy C-means. Instead, we introduce the compositional extension of the distance metrics, according to which we extend the K-means algorithm that is statistically appropriate and correct for grouping the DMs based on their priorities.

In summary, the contributions of this article are as follows:

\begin{itemize}
    \item We show that the proper way to aggregate the priorities is to use the geometric mean method (GMM) and that the arithmetic mean should be avoided. Further, the adaptive weighted geometric mean method (AWGMM) is also developed by using the robust Welsch estimator.
    
     \item The errors of computing the standard deviation of priorities are discussed, and proper Bayesian statistical tests for verifying the difference between the weights of two criteria are developed, the result of which is a probabilistic ranking of criteria. 
     
    \item A proper distance function between two priorities is reviewed, according to which a new clustering method is developed for grouping multiple DMs based on their priorities. 
   
\end{itemize}

The paper is structured as follows. Section \ref{sec:pre} presents the preliminary concepts of MCDM and compositional data, which will be extensively used in the subsequent sections. Section \ref{sec:aggregating dms} is dedicated to the aggregation and the fallacies therein and then puts forward a new robust aggregation method. In Section \ref{sec:std}, we discuss the shortcoming in computing standard deviation and statistical tests in MCDM and develop mathematically sound tests that can be used in such situations. Section \ref{sec:distance} is devoted to the applications of distance-related measures in MCDM, where we also develop some new correct ways which can be used in the context of MCDM. Section \ref{sec:experiment} studies a real-world MCDM example, where we show the use of the proposed methods. Finally, the paper is concluded in Section \ref{sec:conclusion}.

\section{Preliminary}\label{sec:pre}
This section presents the preliminary concepts from MCDM and compositional data, which will be used in the following sections. We first review some rudimentary concepts from pairwise comparison matrices (PCMs) and discuss the consistency of such an MCDM problem. We then look into the compositional data and provide necessary introductory notions.

\subsection{Pairwise comparison matrix}
Pairwise comparison matrices (PCMs) were first introduced and used in AHP (analytic hierarchy process) \citep{ahp, ahp_review}, where they were used to identify the importance of a set of criteria. For a set of criteria $C=\{c_1,c_2,...,c_n\}$, the following definition provides the notion of PCMs.

\begin{Definition}
For a set of \textit{n} criteria, a PCM $M = \{m_{ij}\}_{i,j=1}^n$ is a square matrix of order \textit{n}, whose element $m_{ij}$ expresses the relative importance of criterion $c_i$ over $c_j$.
\end{Definition}

A particular example of a PCM is shown in Table \ref{tab:pcm}. The preferences of criteria over each other are shown on a scale of $1-9$. A number like $8$ in this table indicates that $c_1$ in eight times more important than $c_3$. 

\begin{table}[H]
    \centering
    \caption{An example of pairwise comparison matrix (PCM) for three criteria, i.e., $C=\{c_1,c_2,c_3\}$.}
    \label{tab:pcm}
    \begin{tabular}{c|c c c }
          & $c_1$ & $c_2$ & $c_3$ \\ \hline 
    $c_1$ & 1     & 2     & 8  \\
    $c_2$ & 1/2   & 1     & 4 \\
    $c_3$ & 1/8   & 1/4   & 1 \\
    \end{tabular}
\end{table}

In MCDM, the importance of a set of criteria like $C$ is denoted by a priority vector $w\in \mathcal{R}^n$, where each $w_j$ is non-negative for all $j=1,...,n$, and the sum of all weights come to one. Therefore, each element of a PCM should satisfy the following property:

\begin{align}\label{eq:ahp ratio}
    m_{ij} = \frac{w_i}{w_j}, \quad \forall i,j=1,...,n.
\end{align}

There does not need to exist a unique vector $w$ such that equation \eqref{eq:ahp ratio} satisfies all the elements in a PCM. But if such a unique vector exists, the PCM is said to be fully consistent. The following two definitions provide the notion of consistency for a PCM differently.

\begin{Definition}\citep{ahp}\label{def:full consistent}
A PCM is said to be fully consistent if and only if it satisfies the multiplicative-transitivity property, defined as:

\begin{align}
    m_{ij} = m_{ih} \times m_{hj}, \quad i,j,h=1,...,n.
\end{align}
\end{Definition}

\begin{Definition}\citep{ahp_consistency}
A PCM is said to be fully-consistent if and only if there exists a unique $w \in \mathcal{R}^n$, for which equation \eqref{eq:ahp ratio} holds true.
\end{Definition}

When a PCM is not fully consistent, there are different ways to compute the priorities $w$. The popular methods are the maximal eigenvector method (EVM) and the geometric mean method. The EVM is based on the eigenvalue decomposition, where the criteria priorities are set to be the maximal eigenvector. On the other hand, according to the geometric mean method for PCM, the optimal priorities of criteria are computed by taking geometric means of columns, defined as:

\begin{align}
    w_i = \sqrt[n]{\prod_{j=1}^n \hat{m}_{ij}}, \quad i=1,...,n,
\end{align}

where $\hat{M} = \{\hat{m}\}_{i,j=1}^{n}$ is the column-wise normalized version of a PCM. In both methods, the ratio of weights in equation \eqref{eq:ahp ratio} approximately holds. In the case that the PCM is fully consistent, the column-wise normalization of all columns is the same, making any normalized column the desired criteria priorities.  

\subsection{Compositional data}
The difficulty of processing ratios with common elements in their nominators or denominators is recognized in statistics a long time ago. In his famous paper on spurious correlation in 1897, Karl Pearson pointed out that the correlation of ratios with common parts in their nominators/denominators may not be precisely correct \citep{spurious_correlation}. However, Pearson's precautions went unheeded until the 1980s, when some studies highlighted the methods and tools for analyzing ratios with common elements \citep{compositional_book}. These studies were based on data whose sum is a fixed number, which were called \textit{compositions}. The following definition provides a concise explanation of this notion.

\begin{Definition}\label{composition}\citep{compositional_book}
A composition $w$ of $n$ parts is an $n\times 1$ vector with positive components $w_1,...,w_n$ whose sum is 1. 
\end{Definition}

Be aware that the sum of components of a composition in Definition \ref{composition} is not necessarily $1$ but can be any other number, e.g., 100. Moreover, even the sum of the parts can be unknown and not equivalent for different compositions in the data set. An example of such data is the household budget, where the expenses of families are classified different categories four categories \citep{compositional_book}. Usually, different families have different costs, typically commensurate to their income, but for analyzing such data, the ratio between different expense classes is critical, and the data is thus compositional. In this circumstance, we can divide the value of an expense class for each family by the total expenses of the family to form a compositional vector from the raw expense values. As a result, a different (unknown) constant-sum does not change the nature of the compositional data, where the ratio between different parts matters, not the magnitude of its parts.

Given that compositions have a limiting constraint, the statistical analysis of compositional data must be different. As a concrete example, we cannot assume that a composition follows a normal distribution since it does not lie in the real space $\mathcal{R}^n$. Instead, it lies in lower-dimensional simplices, making the available statistical tools inapplicable to such data. 

The priorities of criteria in MCDM are indeed compositional. As a result,  statistical analysis---even simple arithmetic or geometric mean---should be investigated before applying directly to such data. In the following sections, several statistical methods are analyzed, and proper methodologies for processing priorities in MCDM are developed and put forward, which align with the priorities' compositional nature.

\section{Priorities Aggregation}\label{sec:aggregating dms}

Group members involved in a decision-making problem can be expected to have different preferences, which might be due to different causes, including (i) uncertainty (due to lack of relevant information and proper structuring of the problem); (ii) conflict (due to different values or priorities); or (iii) misunderstanding (due to different perspective and partial information) \citep{belton1997}. There are generally three approaches to handling the differences among group members: sharing, comparing, and aggregating \citep{belton1997}. By sharing, the analyst/facilitator tries to moderate a discussion, for instance, via the decision conferencing \citep{Phillips1990}\citep{mccartt1989}, among the members to address the differences by discussing the causes of different preferences with the ultimate aim of finding common ground and agreement. In this approach, the group is treated as a single DM. In comparing, on the other hand, members are considered individually, and the preferences obtained from members are used for further negotiation and comparison to reach a consensus. Finally, in aggregating, the individual preferences are not negotiated, nonetheless, the analyst/facilitator tries to find common ground where the differences among the preferences are reduced. 

While reaching a consensus among the group members by following sharing or comparing approaches is desirable, a unanimous agreement is not always guaranteed. This is why the aggregating approach could also be seen as complementary to the comparing approach. A responsible analyst/facilitator should try to moderate a discussion among the members and make their views close. However, if reaching a consensus seems impossible, aggregating the preferences (which have already become closer) comes to the picture in the end. Also, in some situations where sharing and comparing is not easy, e.g., when we need to collect the preferences of a relatively large number of participants, aggregating seems the only possible option. Examples could be arriving at a decision in a municipality by collecting the preferences of citizens of a town or formulating a policy by collecting the preferences of electric vehicle users in a region. We refer to \citep{belton2002} for detailed explanations of the three approaches mentioned here. Regarding aggregating the priorities, as we discussed earlier, we need to use operations suitable for priorities, i.e., operations on compositions.

In this section, we first review the compositional approach to aggregating the priorities of multiple DMs and show that the geometric mean of priorities should be utilized for aggregation, and the use of the arithmetic mean should be avoided in this case. Further, a method based on robust statistics is developed for aggregating the priorities that are robust to deviant DMs and identifies the DMs whose priorities are different from the majority of other DMs, making them have a lesser impact on the final aggregated priorities. Identifying deviant DMs can also be used in the negotiation process to converge the preferences of the DMs in the group. The notations used in this article are also shown in Table \ref{tab:notation}.

\begin{table*}[]
    \centering
    \caption{Notation used throughout the paper.}
    \begin{tabular}{c|c}
    variable & description \\ \hline 
         K &  number of DMs\\
         n &  number of criteria \\
         $W \in R^{K\times n}$ & matrix containing all the priorities of $K$ DMs for $n$ criteria \\
         $\hat{W} \in R^{K \times n(n-1)/2}$ & log-ratio transformation of priorities \\ 
         $P$ & performance matrix \\
         $w^g \in \mathcal{R}^n$ & aggregated priority \\
         $\hat{w}\textbf{} \in R^{n(n-1)/2}$ &  log-ratio transformation of the aggregated priority
    \end{tabular}
    \label{tab:notation}
\end{table*}

\subsection{Arithmetic versus geometric mean}
There are two widely-used approaches for aggregating multiple DM priorities: arithmetic and geometric mean. Initially, the arithmetic mean method (AMM) was reported to be the most appropriate method because the geometric mean violates the Pareto optimality \citep{aggregate_1994}. Later on, Forman and Peniwati showed that the geometric mean preserved the Pareto optimality as well \citep{aggregate_1998} and recommended using the geometric mean for aggregating priorities without a solid justification. However, this recommendation is taken for granted, and the arithmetic mean has been used mainly for aggregating the priorities. Even in a more recent review article \citep{ahp_review}, only the arithmetic mean is mentioned as the only means for aggregating the priorities in the AHP.   

These two methods have a fundamental problem of averaging DM priorities, each element showing only the relative importance of one criterion concerning other criteria. Since only the ratio between the weights of different criteria, and not the magnitude of weights, matters, we cannot directly apply the arithmetic or geometric mean over the priorities of multiple DMs. Instead, we first need to compute all the possible ratios between the elements in a priority or weight vector and then take an average over the ratios. For $n$ criteria, all the possible ratios are $n(n-1)/2$ placed in a \textit{compositional average array}. 

\begin{Definition}[Compositional average array \citep{compositional_book}] \label{def:average array}
For an aggregated $n$-part compositions like $w$, the compositional average array is given by 

\begin{equation*}
E = 
\begin{bmatrix}
     0   & \xi_{12} & \xi_{13} & ... & \xi_{1n}  \\
\xi_{21} &    0     & \xi_{23} & ... & \xi_{2n} \\
\vdots & & & & \vdots \\
\xi_{n1} & \xi_{n2} & \xi_{n3} & ... & 0 \\

\end{bmatrix},
\end{equation*}

where $\xi_{ij} = \mathbb{E}\{\ln\frac{w_i}{w_j}\}$, and $\mathbb{E}$ is the mathematical expectation.
\end{Definition}

The mathematical expectation in Definition \ref{def:average array} can be replaced by the average for the empirical analysis since the log-ratios lie within the real space $\mathcal{R}$. Thus, one can compute each element of the compositional average array based on the priorities of $K$ DMs as:

\begin{equation}\label{eq:estimation expectation}
\xi_{ij} = \mathbb{E}\{\ln\frac{w_i}{w_j}\} = \frac{1}{K}\sum_{k=1}^K \ln\frac{W_{ki}}{W_{kj}}, \quad \forall i,j=1,...,n.
\end{equation} 

In addition, it is evident that:

\begin{equation} 
\mathbb{E}\{\xi_{ij}\} = \mathbb{E}\{\xi_{ik}\} + \mathbb{E}\{\xi_{kj}\}, \quad \forall i,j=1,...,n.
\end{equation}

On the other hand, a PCM, like $M$, is said to be fully-consistent if $m_{ij} = m_{ih}\times m_{hj}$ for all $i,j,k$ in the range of the matrix (see Definition \ref{def:full consistent}). Now, if we define the matrix $\hat{E}$ by taking element-wise exponential from $E$, i.e., $\hat{E} = exp(E)$, this matrix can be viewed as a fully-consistent PCM. For fully consistent PCMs, a normalized column would yield the final aggregated priorities. The following lemma shows that a normalized column of matrix $\hat{E}$ is equivalent to the normalized geometric mean of all priorities or the GMM. 

\begin{Lemma}
A normalized column of the exponential-transformed compositional average array is tantamount to the geometric mean method (GMM).  
\end{Lemma}

\begin{Proof}
Without loss of generality, we take the first column of the compositional average array and show that the normalization of the first column is equivalent to the GMM. Let $c \in \mathcal{R}^n$ be the exponential-transformed of the first column, then,
\begin{align}
    c_i &= exp(\xi_{i1}), \quad i=1,...,n, \cr
        &= exp\Bigg(\frac{1}{K} \sum_{k=1}^K \ln \frac{W_{ki}}{W_{k1}}\Bigg)\cr
        &= exp\Bigg(\ln\prod_{k=1}^K \bigg(\frac{W_{ki}}{W_{k1}}\bigg)^{\frac{1}{k}} \Bigg) \cr
        &= \prod_{k=1}^K \Bigg(\frac{W_{ki}}{w_{k1}}\Bigg)^{\frac{1}{K}}.
\end{align}
Now, the normalization of $c$ yields
\begin{align}
    \frac{c_i}{\sum_{j=1}^n c_j} &=  \frac{1}{\sum_{j=1}^n \prod_{k=1}^K \frac{W_{kj}^{\frac{1}{K}}} {W_{k1}^{\frac{1}{K}}} } \prod_{k=1}^K \Bigg(\frac{W_{ki}}{W_{k1}}\Bigg)^{\frac{1}{K}}\cr
    & = \frac{\prod_{k=1}^K W_{ki}^{\frac{1}{K}} }{\sum_{j=1}^n \prod_{k=1}^K W_{kj}^{\frac{1}{K}}},
\end{align}
which is the normalized geometric mean of priorities or the GMM, and that completes the proof.
\end{Proof}

\begin{Corollary}
Taking any average over the raw priorities is not statistically correct. Nevertheless, the analysis based on compositional data showed that the GMM is the proper average of all the priorities. This implies that using the arithmetic mean for aggregating DM priorities should be avoided.   
\end{Corollary}

\subsection{Robust aggregation based on the Welsch estimator}

In equation \eqref{eq:estimation expectation}, the mathematical expectation is replaced by the averaging over the available samples, i.e., the priorities of $K$ DMs, and the final aggregated outcome becomes equivalent to the GMM. Another option is to use more robust statistics, such as the median, to produce more robust priorities for deviant DMs. In aggregating priorities, a deviant is a DM whose preferences deviate significantly from most other DMs. Note also that using the median on the original data is erroneous. The final aggregated results using the median in equation \eqref{eq:estimation expectation} are not tantamount to the GMM but are a more robust surrogate. However, the median would not lead to a fully consistent PCM based on the associated compositional average array.

Another choice, instead of averaging, is to use different estimators. In robust statistics, there are a number of estimators that can provide more robust estimations. Such estimators are called M-estimators that can replace the mean or median in equation \eqref{eq:estimation expectation}. For example, the Welsch M-estimator is one of the well-known estimators that has shown promising performance in noisy environments \citep{entropy_cnv, gene_correntropy}. To use this estimator for computing the compositional average array, we consider the log-ratio transformed data $\hat{W}$ and estimate the aggregated log-ratio transformed priority $\hat{w}^g$ by solving the following optimization problem:
\begin{align}\label{eq:robust aggregation}
    \min_{\hat{w}^g} \sum_{k=1}^K \phi(\Vert \hat{W}_{k\bigcdot} - \hat{w}^g\Vert),
\end{align}

where $\phi(x) = exp(-x^2/\sigma^2)$ is the Welsch estimator. For $\phi(x) = x$, minimization \eqref{eq:robust aggregation} yields the same solution as the arithmetic mean of log-ratio transformed data, so the result of the compositional average array would be identical to the GMM. Using the Welsch, a more robust estimator, we expect that the aggregation outcome will be more robust to deviants, i.e., the DMs whose preferences are significantly different from the majority of other DMs. Problem \eqref{eq:robust aggregation} is not convex \citep{hq_multiplicative}, but the following lemma paves the way for an efficient solution.

\begin{Lemma}[\cite{hq_multiplicative}]\label{lemma:hq}
For a fixed $x$, there is a potential dual function $\psi$ such that:
\begin{equation}
    \phi(x_i) = \inf_{\alpha_i} \alpha_i x_i^2 + \psi(\alpha_i),
\end{equation}
where $\psi(.)$ is the convex conjugate of $\phi(.)$, and  $\alpha_i > 0$ is an auxiliary variable, which is determined by the so-called minimizer function $\delta(.)$ defined as: 
\begin{equation}
    \delta(x) = exp(-\frac{x^2}{\sigma^2}).
\end{equation}
\end{Lemma}

Using Lemma \ref{lemma:hq}, problem \eqref{eq:robust aggregation} can be written as:

\begin{equation}
    \min_{\hat{w}^g,\alpha} \sum_{k=1}^K \alpha_k \Vert \hat{W}_{k\bigcdot} - \hat{w}^g \Vert^2 + \psi(\alpha_k).
\end{equation}

Thus, the following steps must be iterated until convergence is reached:

\begin{align}\label{eq:HQ iteration}
    &\alpha_k = \delta\bigg(\Vert \hat{W}_{k\bigcdot} - \hat{w}^g\Vert\bigg) = exp\Bigg(-\frac{\Vert \hat{W}_{k\bigcdot} - \hat{w}^g\Vert}{\sigma^2}\Bigg),\quad \forall k=1,...,K, \nonumber \\
    &\hat{w}^g = arg\min_{\hat{w}^g} \sum_{k=1}^K \alpha_k \Vert \hat{W}_{k\bigcdot} - \hat{w}^g\Vert^2.
\end{align}

For the second step in \eqref{eq:HQ iteration}, we need to take the derivative and find the optimal solution as:
\begin{align}
    &\frac{\partial}{\partial \hat{w}^g}\sum_{k=1}^K \alpha_k \Vert \hat{W}_{k\bigcdot} - \hat{w}^g\Vert_2^2 = 0 \cr
    &\Rightarrow\sum_{k=1}^K \alpha_k \hat{W}_{k\bigcdot} = \sum_{k=1}^K \alpha_k \hat{w}^g \cr 
    &\Rightarrow \hat{w}^g = \sum \lambda_k \hat{W}_{k\bigcdot}, \quad \lambda_k = \frac{\alpha_k}{\sum_{j=1}^K \alpha_j}.
\end{align}

In addition, the performance of the Welsch estimator is heavily reliant on selecting its parameter $\sigma$. As the recent studies suggest \citep{hq_cnv}, this parameter can be recursively updated in each iteration as:

\begin{align}
    \sigma = \frac{\sum_{k=1}^K \Vert \hat{W}_{k\bigcdot} - \hat{w}^g\Vert_2^2}{n^2}.
\end{align}

Algorithm \ref{algo:aggregate} summarizes the overall procedure for aggregating the priorities of multiple DMs by using the Welsch estimator. 

\begin{algorithm}[H]
\caption{Adaptive weighted geometric mean method (AWGMM)}
\label{algo:aggregate}
\begin{algorithmic}
\State \textbf{Input:} Priorities $W\in R^{K\times n}$.
\While{\textit{NotConverged }}
    \State $\alpha_k = exp(-\frac{\Vert \hat{W}_{k\bigcdot} - \hat{w}^g\Vert}{\sigma^2}),\quad \forall k=1,...,K$
    \State $\lambda_k = \alpha_k / \sum_j \alpha_j$,  \quad $k=1,2,...,K$
    \State $\hat{w}^g = \sum_k \lambda_m \hat{W}_{k\bigcdot}$  
\EndWhile
\State \textbf{Output} Final weight ratios $\hat{w}^g$, $\lambda$
\end{algorithmic}
\end{algorithm}

The value of $\lambda_k$ in Algorithm \ref{algo:aggregate} shows the contribution of each DM to the final aggregated priorities, and the DMs with deviating preferences from the majority of other DMs are assigned a lower $\lambda_k$. As a result, the DMs with deviating preferences from the vast majority of other DMs can be detected by considering their associated $\lambda_k$. Note also that $\lambda \in R^K$ satisfies the non-negativity and unit-sum constraints, allowing us to view it as a weight for DMs based on their opinion proximity to other DMs.     

\begin{Remark}
The values of $\lambda$ from Algorithm \ref{algo:aggregate} identifies the deviant DMs. Instead of aggregation, the decision facilitator/analyst can use these values to understand what DMs have different preferences and use such information in the negotiation process. 
\end{Remark}

\begin{Remark}
We do not claim the preference of $\ w^g_{AWGMM}$ and  $ w^g_{GMM}$ over each other. While both are mathematically sound approaches for aggregating priorities, their main difference is handling the deviants. While  $ w^g_{GMM}$ assigns equal weights to all the DMs,  $\ w^g_{AWGMM}$ assigns lower weights to the DMs whose priorities are far from the majority of the DMs. We think the suitability of the two approaches depends on a particular decision-making situation; if all the DMs involved in the decision-making process must have equal contributions, then the GMM must be used. However, if the majority opinion is of more interest, then AWGMM seems more appropriate. In any case, using the arithmetic mean for aggregating priorities should be avoided.
\end{Remark}

After obtaining $\hat{w}^g$ from Algorithm \ref{algo:aggregate}, it can be placed in a compositional average array, which happens to be fully consistent, making the aggregated priorities be obtained by normalizing a column of such a matrix. The following lemma proves the consistency of the compositional average array built based on $\hat{w^g}$.

\begin{Lemma}\label{lm:awgmm full consistent}
The compositional average array computed based on Algorithm \ref{algo:aggregate} is fully consistent, thereby providing unique aggregated priorities.
\end{Lemma}

\begin{Proof}
Algorithm \ref{algo:aggregate} outputs the log-ratio transformed $\hat{w}^g$ and $\lambda \in R^{K}$. Accordingly, the compositional average array can be constructed as in Definition \ref{def:average array}, where 
\begin{align*}
    \xi_{ij} &= \mathbb{E}\bigg\{\ln\frac{W_{ki}}{W_{kj}}\bigg\} \\
             &= \sum_{k=1}^K \lambda_k \ln\frac{W_{ki}}{W_{kj}}, \quad \forall i,j=1,..., n,
\end{align*}
where $\lambda$ is obtained from Algorithm \ref{algo:aggregate}. It is now simple to show that:
\begin{align*}
    \xi_{ij} = \xi_{ik} + \xi_{kj},
\end{align*}
which means that the exponential-transformed of such a matrix is fully consistent and provides unique aggregated priorities, and that completes the proof.
\end{Proof}

The proposed aggregation method has several features, some of which are listed in the following:

\begin{itemize}
    \item \textbf{Pareto optimality} entails that if all DMs in a group prefer \textit{A} over \textit{B}, then the group decision should also favor \textit{A} \cite{aggregate_1998}. In the proposed aggregation method, if all DMs favor criterion $i$ over $j$, then:
    \begin{align}
        \ln\frac{W_{ki}}{W_{kj}} > 0, \quad \forall k=1,..,K,
    \end{align}
    and since $\lambda_k$ is non-negative and is summed to one in Algorithm \ref{algo:aggregate}, it follows 
    \begin{align}
        \sum_{k=1}^K \lambda_k \ln\frac{W_{ki}}{W_{kj}} > 0,
    \end{align}
    
    which means that the proposed aggregation method satisfies Pareto optimality.
    
    \item \textbf{Non-dictatorship} refers to the fact that no individual priorities become the priorities of the group automatically, regardless of the preferences of other members in the group. The proposed method assigns a weight to each DM, the magnitude representing the corresponding DM's contribution to final aggregated weights. Although the procedure might give some DMs a lower weight (even nearly zero), the weights are only assigned after considering the priorities of all DMs in a group. In addition, if a DM is added or removed from a group, the weights of DMs and the final aggregated priorities will change, confirming that the aggregated priorities are not automatically biased toward one of the DMs. 
    
    \item \textbf{Recognition} means that the group decision is arrived at after considering all the members' priorities. In the proposed aggregation method, each DM is assigned a weight after considering the priorities of all group members: If the priorities of a DM are different from those of other group members, then it is assigned a lower weight. It means that the final aggregated priorities are computed based on the priorities of \textit{all} decision-makers. Further, adding or removing a DM will change the weights assigned to each DM and the final aggregated priorities, which also corroborates that AWGMM considers all DMs' priorities before arriving at the aggregation priorities.
\end{itemize}

\subsection{An illustrative example}\label{sec:aggregating example}
This section illustrates the aggregation procedure through an example. In this regard, assume that five DMs have expressed their preferences on four criteria $C=\{c_1,c_2,c_3,c_4\}$, and the matrix $W$ containing the priorities of five DMs is as follows:

\begin{align}
    W =  
    \begin{blockarray}{ccccc}
    & c_1 & c_2 & c_3 & c_4 \\
 \begin{block}{c(cccc)}   
DM1 &  0.220 &  0.435  &  0.295  &  0.050 \\
DM2 &  0.210 &  0.434  &  0.312  &  0.044 \\
DM3 &  0.363 &  0.312  &  0.107  &  0.218 \\
DM4 &  0.243 &  0.386  &  0.332  &  0.039 \\
DM5 &  0.227 &  0.381  &  0.339  &  0.053 \\
\end{block}
\end{blockarray}.
\end{align}

The result of the arithmetic mean is: 

\begin{align} \label{eq:example amm}
w^g_{AMM} = \begin{bmatrix}
0.253 & 0.389 & 0.277 & 0.081
\end{bmatrix}.
\end{align}

We first create the compositional average array, as defined in Definition \ref{def:average array}. In this regard, we compute element $\xi_{12}$ as:

{\small
\begin{align*}
    \xi_{12} &= \frac{1}{5} \Bigg(\ln\frac{0.220}{0.435} + \ln\frac{0.210}{0.434} + \ln\frac{0.363}{0.312} + \ln\frac{0.243}{0.386} + \ln\frac{0.227}{0.381}\Bigg) \cr
    &=  -0.446. 
\end{align*}
}

Similarly, other elements in the compositional average array are computed, the result of which is as follows:

\begin{align}
       E = \begin{pmatrix}
    0       & -0.446    &   -0.036  &  1.371 \\
    0.446   & 0         &    0.411  &  1.817 \\
    0.036   & -0.411    &    0      &  1.407 \\
   -1.371   & -1.817    &   -1.407  &    0
     \end{pmatrix}.
\end{align}

By taking exponential and normalizing a column of this matrix, we arrive at the aggregated priorities as:

\begin{align} \label{eq:example gmm}
w^g_{GMM} = \begin{bmatrix}
0.260 & 0.405 & 0.269 & 0.066
\end{bmatrix},
\end{align}

which is identical to the GMM and is considerably different from \eqref{eq:example amm}. We further apply Algorithm \ref{algo:aggregate} to find the aggregated priorities. This algorithm works with the log-ratio transformed data $\hat{W}$ and has, as one of its outputs, $\lambda \in R^K$ that works as a weight for different DMs. The $\hat{W}$ and $\lambda$ for this example are:

\[\small
 \hat{W} =  \begin{blockarray}{ccccccc}
    \lambda & \frac{c_1}{c_2} & \frac{c_1}{c_3} & \frac{c_1}{c_4} & \frac{c_2}{c_3} & \frac{c_2}{c_4} & \frac{c_3}{c_4} \\
    \begin{block}{c(cccccc)} 
0.31 & -0.681  &  -0.294  & 1.480 & 0.387 & 2.161  &  1.774 \\
0.33 & -0.722  &  -0.391  & 1.560 & 0.331 & 2.282  &  1.952 \\
0.00 &  0.151  &   0.221  & 0.511 & 1.070 & 0.360  & -0.710 \\
0.24 & -0.519  &  -0.403  & 1.461 & 0.116 & 1.980  &  1.864 \\
0.12 & -0.462  &  -0.311  & 1.841 & 0.151 & 2.303  &  2.152  \\
\end{block}
\end{blockarray}.
\]

As is evident from this matrix, the third DM is assigned a weight of approximately zero\footnote{Note that it is not absolute zero, but an infinitesimal number.}, mainly because their preferences are significantly different from others. Thus, it is treated like a deviant and does not influence the final aggregated log-ratio $\hat{w}^g$ that is computed as:

\begin{align*}
    \hat{w}^g =  \begin{bmatrix}
    -0.628 & -0.354 & 1.546 & 0.274 & 2.174 & 1.90
    \end{bmatrix},
\end{align*}

which can be placed in an array as:

\begin{align}
    E = \begin{pmatrix}
     0     & -0.628  &   -0.354  &  1.546 \\
    0.628  &  0      &    0.274  &  2.174 \\
    0.354  & -0.274  &    0      &  1.90 \\
   -1.546  & 2.174    &   -1.90   &    0    
   \end{pmatrix}.
\end{align}

It can be verified that the exponential-transformed of this matrix is a fully consistent PCM, which was proved in Lemma \ref{lm:awgmm full consistent}. Now, if we take the exponential and then normalize a column, the final aggregated priorities are obtained as:
\begin{align}\label{eq:example awgmm}
    w^g_{AWGMM} = 
    \begin{bmatrix}
        0.225 & 0.410 & 0.319 & 0.046
     \end{bmatrix}.
\end{align}

As expected, the average computed by the AWGMM in \eqref{eq:example awgmm} is different from those in \eqref{eq:example gmm} and \eqref{eq:example amm} since one of the DMs is assigned a small weight (nearly zero) and has therefore minimal impact on the final aggregated priorities. Note also that the values of $\lambda$ can be used in the negotiation, as the third DM has significantly different preferences than others. 

\section{Standard deviation and statistical tests}\label{sec:std}
Similar to the discussion regarding the central tendency of priorities, the standard deviation for the priorities is defined differently. The change of central tendency and standard deviation influences statistical tests, such as paired \textit{t}-test and Wilcoxon Signed-rank test, which are often used in processing the priorities. The use of such statistical tests is to verify if the difference between the weights of the two criteria is significantly different \citep{fal_statest_1}. In this section, we first study the estimation of standard deviation from the perspective of compositional data analysis and show that such a definition would provide correct and more meaningful results than computing the standard deviation from raw priorities. We then combine compositional data analysis with Bayesian statistics to calculate the probability that one criterion is more important than another based on the priorities of multiple DMs and provide a probabilistic ranking of the criteria.

\subsection{Standard deviation}
The standard deviation of elements in priorities has also been used without considering the inherent constraints in the compositional data \citep{fal_standardDeviation_1,fal_standardDeviation_2}. Like the arithmetic mean, the standard deviation calculation based on raw priorities is also incorrect and invalidates the consequent decisions and methods. In addition, the standard deviation estimated based on the priorities typically has a magnitude that is as large as or even larger than the mean of the priorities (see \citep{fal_standardDeviation_1, fal_standardDeviation_2}, thereby having narrow applicability in practice.

For compositional data, we need to compute the deviation of all possible ratios in a composition. Therefore, we define the \textit{compositional deviation array} that includes the standard deviation of all possible ratios. 

\begin{Definition}[Compositional deviation array \citep{compositional_book}]\label{def:deviation array}
For an $n$-part composition like $w$, the compositional deviation array is given by 

\begin{equation*}
\mathcal{T} = 
\begin{bmatrix}
     -   & \tau_{12} & \tau_{13} & ... & \tau_{1n}  \\
\tau_{21} &    -     & \tau_{23} & ... & \tau_{2n} \\
\vdots & & & & \vdots \\
\tau_{n1} & \tau_{n2} & \tau_{n3} & ... & - \\

\end{bmatrix},
\end{equation*}

where $\tau_{ij} = \sqrt{var\big(\ln\frac{w_i}{w_j}\big)}$, and $var$ is the variance operator.
\end{Definition}

The variance in definition \ref{def:deviation array} can be replaced by the empirical variance defined as:
\begin{equation}\label{eq:var}
\tau_{ij}^2 = var\big(\ln\frac{w_i}{w_j}\big) = \frac{1}{K-1}\sum_{k=1}^K \bigg(\ln\frac{W_{ki}}{W_{kj}}-\xi_{ij}\bigg)^2, \quad \forall i,j=1,...,n.
\end{equation} 

Definition \ref{def:deviation array} shows that for $n$ criteria, we indeed have $n(n-1)/2$ unique standard deviations, equivalent to the number of possible ratios for an $n$-part composition. This means that, based on such a definition, the analysis based on the average and standard deviation of each criterion computed according to raw priorities makes no sense. Besides, by using the deviation and average arrays, it is possible to provide more insights regarding the priorities of different DMs over a set of criteria, which provides more description of DMs' priorities and the relative importance of different criteria rather than a sole ratio. 

Before illustrating the usefulness of the compositional deviation array through an example, it is worth noting that the definition of variance for the deviation array can also be replaced by the median absolute deviation (MAD), which is based on the median and is a more robust surrogate for standard deviation. In addition, based on the proposed aggregation method AWGMM, we can define a robust variance as:

\begin{align}
    var\bigg(\ln\frac{w_i}{w_j}\bigg)^2 = \sum_{k=1}^K \lambda_k \bigg(\ln\frac{W_{ki}}{W_{kj}} - \xi^{AWGMM}_{ij}\bigg)^2,
\end{align}
where $\lambda$ and $\xi^{AWGMM}$ are computed based on the proposed aggregated method in Algorithm \ref{algo:aggregate}. Given these definitions, we illustrate the use of standard deviation by an example. In addition, since the magnitudes of values in average and deviation arrays are identical, we can place the numbers in an array to summarize the priorities of multiple DMs in one array only. The next definition presents such an array.

\begin{Definition}[Compositional average-deviation array]\label{def:avg-var}
For an n-part composition like \textit{w}, the compositional average-deviation array (AD) is defined as:

\begin{equation*}
\begin{pmatrix}
     0   & \xi_{12} & \xi_{13} & ... & \xi_{1n}  \\
\tau_{21} &    0     & \xi_{23} & ... & \xi_{2n} \\
\vdots & & & & \vdots \\
\tau_{n1} & \tau_{n2} & \tau_{n,3} & ... & 0 \\

\end{pmatrix},
\end{equation*}
where $\xi_{ij}$ is the average, i.e., mean, median, or a robust estimation based on Algorithm \ref{algo:aggregate}, and $\tau_{ji}$ is the associated standard deviation to $\xi_{ij}$.
\end{Definition}

\begin{Example}
We compute the compositional average-deviation array for the example in Section \ref{sec:aggregating example} by using the mean, median, and the method proposed in Algorithm \ref{algo:aggregate}. We first begin with the mean and standard deviation, that is computed as:

\begin{align}
  AD_{mean} = 
    \begin{blockarray}{ccccc}
    & c_1 & c_2 & c_3 & c_4 \\
 \begin{block}{c(cccc)}   
c1 & 0      &  -0.446 &  -0.035  &  1.370 \\
c2 & 0.351  &       0 &   0.410  &  1.817 \\
c3 & 0.704  &   0.386 &       0  &  1.406 \\
c4 & 0.504  &   0.824 &   1.191  &      0 \\
\end{block}
\end{blockarray}.
\end{align}

The magnitude of the above-diagonal values suggests the difference between the two criteria, and its sign indicates which criterion is more important. For instance, the value $-0.446$ indicates that criterion $c_2$ is more important than criterion $c_1$, since the value is negative. The biggest difference is between $c_2$ and $c_4$ whose magnitude of average differences is $1.817$. The lower-diagonal entries represent the standard deviation, which can help realize how reliable the difference between the two criteria is based on a group of DMs' priorities. For instance, the average difference between $c_1$ and $c_3$ has the infinitesimal value of $-0.035$, while the standard deviation is $0.704$. Therefore, one can readily understand that these criteria have similar priorities or importance to the group of DMs. 

The same average-variation array can be computed by using the median and median absolute deviation, as well as the proposed robust aggregation of priorities, i.e., 

\begin{align}
       AD_{median} = \begin{pmatrix}
 0      &  -0.518  &  -0.311  &   1.479 \\
0.162  &       0  &   0.330  &   2.160 \\
0.080  &   0.179  &       0  &   1.864 \\
0.081  &   0.142  &   0.090  &        0\\
     \end{pmatrix},
\end{align}

\begin{align}
       AD_{AWGMM} = \begin{pmatrix}
    0  & -0.628 &  -0.354  &  1.546  \\
0.100  &      0 &   0.274  &  2.175  \\
0.048  &  0.113 &       0  &  1.901  \\
0.117  &  0.122 &   0.118  &      0  \\
     \end{pmatrix}.
\end{align}

The matrices computed based on the median and the proposed aggregated method are similar but significantly different from those computed based on the mean. As an instance, the difference between criteria $c_1$ and $c_3$ are more significant than that computed by mean ($-0.311$ and $-0.354$ for median and AWGMM, respectively, but $-0.035$ for mean), while the standard deviation is also tiny ($0.080$ and $0.048$ for median and AWGMM, respectively, but $0.704$ for mean), making the difference between the two criteria significantly different. By looking at the priorities provided by DMs, we can readily realize that $c_3$ is consistently better than $c_1$, except for the third DM, and the difference between the weights of these two criteria for four DMs is also significant with the average ratio of $1.42$. At the same time, the third DM favors $c_1$ to $c_3$ with the factor of $3.39$. This influences the average differences between the two criteria and skews the mean statistics. However, in more robust approaches, such as median and AWGMM, such priorities have less of an impact on the aggregated statistics, allowing us to capture robustly the statistical description of criteria importance. The same considerable difference exists for comparing $c_1$ and $c_4$ as well.
\end{Example}

\subsection{Statistical comparison of two criteria: A Bayesian approach}
In the previous section, we show how notions of compositional data can help describe the importance of criteria in MCDM, given a number of priorities from multiple DMs. Along the same line, this section is devoted to studying the statistical tests for comparing the significance of the difference between the criteria based on various DMs' priorities.

We have three statistical tests for comparing two criteria: paired $t$-test (or one-sample $t$-test), Wilcoxon Signed-rank test, and Sign test. There is a comprehensive comparison of these tests in practical problems \citep{demsar,om_statistical}, concluding that each test is appropriate in a given circumstance. However, the statistical tests based on $p$-value have many drawbacks \citep{benavoli_allBayesian}, making the outcome of the tests unreliable and of little practical importance. Therefore, it is highly recommended to use Bayesian tests instead of using $p$-value inferences. The use of Bayesian statistics also allows us to make a more meaningful comparison: We can compute the extent to which one criterion is more important than another based on the priorities of a group of DMs. Such a meaningful comparison was the primary driver of some recent studies in group MCDM, which tried to compute the extent to which one criterion is more important than another by using the standard deviation of priorities \citep{fal_standardDeviation_1,fal_standardDeviation_2}. While these studies are fallacious due to an incorrect and improper calculation of standard deviation, we here provide a statistically-sound method based on compositional data and Bayesian statistics. We first review two basic definitions introduced in \citep{bayesian_bwm}.

\begin{Definition}[Credal Ordering \citep{bayesian_bwm}]
For a pair of criteria $c_i$ and $c_j$, the credal ordering $O$ is defined as:
\begin{align}
O = (c_i,c_j,R,d),
\end{align}
where
\begin{itemize}
\item R is the relation between the criteria $c_i$ and $c_j$, i.e., $<$, $>$, or $=$;
\item $d\in [0,1]$ represents the confidence of the relation.
\end{itemize}
\end{Definition}

\begin{Definition}[Credal Ranking \citep{bayesian_bwm}]
For a set of criteria $C = (c_1,c_2,...,c_n)$, the credal ranking is a set of credal orderings which includes all pairs $(c_i,c_j)$, for all $c_i,c_j \in C$. 
\end{Definition}

Using Bayesian statistics to compare every pair of criteria in a given problem will finally result in the credal ranking of all criteria. What is required to be computed is the confidence $d$ for each credal ordering, that could be computed by the Bayesian counterpart of three tests: paired \textit{t}-test, Sign test, and Wilcoxon Signed-rank test.  

The paired \textit{t}-test requires the average and the standard deviation of differences, which can be simply supplied by using the average-deviation array. One of the three arrays, i.e., mean, median, and AWGMM, can be used to conduct the paired\textit{ t}-test. There are a number of Bayesian counterparts for the sample paired \textit{t}-test \citep{bayesian_t1,bayesian_t2,bayesian_t3}. For the case of comparing criteria importance, the Bayesian test proposed in \citep{bayesian_t1} is recommended since it takes the average, standard deviation, as well as the sample size (i.e., the number of DMs in our case) and provides the extent to which one criterion is more important than another, allowing us to experiment with the average-deviation arrays.

For the Sign test, we need to count the DMs that favor one criterion over another and then use the beta-binomial conjugate as a Bayesian test, according to which we can compute the confidence in the credal orderings.

However, developing a Wilcoxon Signed-rank test for compositional data is a bit tricky. Suppose we apply the Wilcoxon Signed-rank test directly to the priorities of multiple DMs. In that case, it accounts for the difference between the weights of two criteria for all the DMs, assigns a rank based on the magnitude of the difference, and computes the statistics. However, for compositional data like the priorities of DMs, the ratio between the weights for each DM should be considered. Therefore, instead of computing the difference between the weights of the two criteria, we should calculate the \textit{ratio} and assign a rank based on the magnitude of the ratios. So, if two criteria are deemed the same, the ratio between their weights should be one. Instead of taking the ratios and comparing them against one, we can take the logarithm of the weights and then apply the conventional Wilcoxon Signed-rank test: If two criteria have the exact weights, then the log-ratio will be zero, and the deviation of the log-ratio from zero indicates that one criterion is significantly more important than another.

For $K$ DMs, the proposed Wilcoxon-type test contains the following steps:

\begin{itemize}
    \item [Step 1.] Compute the ratio $r_k$ between the weights of the two criteria for all the DMs. 
    
    \item [Step 2.] Compute $\hat{r}_k$ by taking the logarithm of ratios $r_k$. Then, rank $\hat{r}_k$ according to their absolute magnitude. In the case of ties, the average rank is assigned.
    
    
    \item [Step 3.] Compute $R^+$ as the sum of ranks of the DMs whose corresponding $\hat{r}_k$ is positive.
    
    \item [Step 4.] Similarly, $R^-$ is computed as the sum of ranks of DMs whose corresponding $\hat{r}_k$ is negative.
    
    \item [Step 6.] For \textit{p}-value statistics, define $T=min(R^+,R^-)$. Most statistical books contain a table of exact critical values for $T$ and $K$. For the Bayesian test, the ranks are given to the Bayesian Wilcoxon test \citep{bayesian_wilcoxon} and compute the extent to which one criterion is more important than another.

\end{itemize}

\begin{table}[]
    \centering
    \caption{An example of applying the proposed Wilcoxon-type test for verifying the significance of the difference between two criteria based on the priorities of multiple DMs.}

    \begin{tabular}{c|c c | c c  c }
    & $c_1$ & $c_2$ & log-ratio & rank \\ \hline 
DM1  & 0.125	&  0.243  &  -0.6650 &  13 \\
DM2  & 0.143	&  0.224  &  -0.4490 &  9  \\
DM3  & 0.147	&  0.231  &  -0.4520 &  10 \\
DM4  & 0.164	&  0.209  &  -0.2420 &  6  \\
DM5  & 0.197	&  0.151  &   0.2660 &  7  \\
DM6  & 0.157	&  0.256  &  -0.4890 &  12 \\
DM7  & 0.153	&  0.232  &  -0.4160 &  8  \\
DM8  & 0.115	&  0.249  &  -0.7730 &  14 \\
DM9  & 0.178	&  0.167  &   0.0640 &  1  \\
DM10 & 0.164	&  0.183  &  -0.1100 &  2  \\
DM11 & 0.175	&  0.211  &  -0.1870 &  5  \\
DM12 & 0.168	&  0.192  &  -0.1340 &  3  \\
DM13 & 0.155	&  0.251  &  -0.4820 &  11 \\
DM14 & 0.126	&  0.273  &  -0.7730 &  15 \\
DM15 & 0.199	&  0.17	  &   0.1580 &  4  \\
    \end{tabular}
    \label{tab:wilcoxon criteria}
\end{table}

We explain the procedure of the Wilcoxon-type problem using an example. Table \ref{tab:wilcoxon criteria} shows the weights of two criteria obtained based on the preferences of 15 DMs. Each row in this table corresponds to each DM. In the last two columns, we show the log ratios and the associated rank of each DM based on the Wilcoxon Signed-rank test. Accordingly, $R^+$ and $R^-$ are calculated as:

\begin{align}
    &R^+ = 7 + 1 + 4 = 12 \cr
    &R^- = 108.
\end{align}

Given $R^+$ and $R^-$, we can then apply the Bayesian Wilcoxon Signed-rank test to compute the extent to which one criterion is more important than another based on the priorities of a group of DMs.

\section{Distance metrics and clustering methods}\label{sec:distance}
In processing multiple DM priorities, the Euclidean distance is widely used, where it directly computes the distance based on the original priorities. In compositional data analysis, however, the distance between two compositions is defined differently, taking the nature of the compositions into account. In this section, the Aitchison distance \citep{aitchison_distance}, which is a proper and arguably the most popular distance metric for compositional data, is reviewed, according to which a clustering method is developed to cluster multiple DMs based on their priorities. 

\subsection{Distance metrics for priorities in group MCDM}
The Euclidean distance is typically used as the distance measure between two priorities. Let $w, w' \in \mathcal{R}^n$ be the priorities of two DMs, the Euclidean distance is defined as:

\begin{align}
    d_e(w,w') = \sqrt{\sum_{i=1}^n (w_i - w'_i)^2}.
\end{align}

The Euclidean distance is based on the Euclidean space and computes the distance based on the original priorities that lie on a simplex (not Euclidean space). Therefore, a proper distance metric should be used to measure the distance more in line with the compositional nature of the priorities. The Aitchison distance is arguably the most popular distance metric in compositional data analysis, which also correlates with the Euclidean distance. The Aitchison distance is defined as the Euclidean distance of the log-ratio transformed data \citep{aitchison_distance}, i.e.,

\begin{align}\label{eq:aitchison dist}
    d_a(w,w') = \sqrt{\sum_{i=1}^{n-1} \sum_{j=i+1}^n \Bigg(\ln\frac{w_i}{w_j} - \ln\frac{w'_i}{w'_j}\Bigg)^2}.
\end{align}

Similarly, the mean absolute deviation (MAD) distance for the compositional data, shown by \textit{MADC}, can be defined as:

\begin{align}\label{eq:maec}
    d_{MADC}(w,w') = \sum_{i=1}^{n-1} \sum_{j=i+1}^n \bigg\vert \ln\frac{w_i}{w_j} - \ln\frac{w'_i}{w'_j}\bigg\vert.
\end{align}

As a result, the distance metrics in \eqref{eq:aitchison dist} or \eqref{eq:maec} should be used for clustering the DMs based on their priorities, which respect the compositional nature of the priorities. 

\subsection{K-means clustering for priorities}
Grouping DMs into a number of clusters is also used in group MCDM \citep{fal_clustering_1}\citep{fal_clustering_2}. One way is to group them based on their priorities using clustering methods, the core building block of which is a distance function. Using an improper distance function would group the DMs on a wrong basis. On top of that, the centroids of clusters would not necessarily satisfy the unit-sum constraint, thereby failing to represent the priorities properly. To prevent these pitfalls, we now extend the K-means clustering algorithm that uses a compositional distance metric, e.g., distance metrics in \eqref{eq:aitchison dist} or \eqref{eq:maec}, to group the DMs based on their priorities.

The K-means needs to know the number of clusters, $o$, and identifies $o$ centroids of clusters $l_1,...,l_o$. The standard K-means uses the Euclidean distance and the arithmetic mean for clustering. However, a compositional distance metric and normalized geometric mean should be utilized for clustering compositional data. 

The steps required to cluster compositional data $W_i, i=1,..., K$ by using \textit{K}-means are as follows:

\begin{itemize}
    \item [Step 1] Place centroids $l_1,...,l_o$ at random locations.
    \item [Step 2] Until convergence, repeat the following steps:
    \begin{itemize}
    \item [Step 2.1] For each $W_i$, find the nearest centred $l_j$ as:
        \begin{align}
            j = arg\min_j d_a(w_i,l_j),
        \end{align}
    where $d_e(.,.)$ is a compositional distance. Then, assign $W_i$ to cluster $j$.
    
    \item [Step 2.2]  For each cluster $j=1,...,o$, update the centroids as the mean of the points within the cluster, i.e.,
        \begin{align}
         l_j =
        \frac{\prod_{k=1}^{K_j} W_{k.}^{\frac{1}{K_j}} }{e^T \bigg(\prod_{k=1}^{K_j} W_{k.}^{\frac{1}{K_j}}\bigg)},
        \end{align}
        where $K_j$ is the number of priorities in cluster $j$, $\prod$ and power are element-wise operations, and $e$ is a vector with elements of one.
    \end{itemize}
\end{itemize}

Since the distance function and the average are different from those of the standard \textit{K}-means, the outcome of clustering will be distinct as well. In addition, the centroids identified by the proposed clustering method would be compositional, while the centroids of standard \textit{K}-means for clustering compositional data are not necessarily compositional, e.g., the centroids are not compositional for the compositional mean absolute deviation.

\section{Numerical Results}\label{sec:experiment}
In this section, we show the correct ways of analyzing the priorities of multiple DMs through a real case study in airline baggage handling. The baggage handling system is essential to ground handling operations and impacts passengers' satisfaction. A recent study develops a model, namely SERVQUAL, for assessing the quality of service for baggage handling \cite{servqual}. Grounded on the literature review and the interviews with the passengers, the SERVQUAL includes five main criteria for evaluating the quality of the airline baggage handling systems: \textit{tangibles}, \textit{reliability}, \textit{responsiveness}, \textit{assurance}, and \textit{empathy}. The estimate the importance of the criteria, the preferences of 148 passengers from several nationalities are elicited according to the best-worst method \cite{bayesian_bwm}. The analyses conducted in this section are based on the priorities of the 148 participants. The data and MATLAB implementation of the corresponding analyses are publicly available\footnote{\href{https://github.com/Majeed7/MCDMfallacies_compositional}{https://github.com/Majeed7/MCDMfallacies\_compositional}}.

\subsection{Aggregating priorities}
We first look into the aggregation of priorities by different methods. We first compute by the AMM (which is not correct but is typically used in the literature), and the result is:

\begin{align} \label{eq:exp amm}
w^g_{AMM} = \begin{bmatrix}
0.1397 & 0.3459 & 0.2289 & 0.1519 & 0.1336
\end{bmatrix}.
\end{align}

Similarly, the results of the GMM and AWGMM are as follows:
\begin{align} \label{eq:exp gmm}
w^g_{GMM} = \begin{bmatrix}
0.1376 & 0.3502 & 0.2347 & 0.1527 & 0.1248
\end{bmatrix}, \cr
w^g_{AWGMM} = \begin{bmatrix}
0.1234 & 0.4462 & 0.1932 & 0.1490 & 0.0883
\end{bmatrix}.
\end{align}

The AMM has a different aggregation than the GMM, so the result of such aggregation can distort the follow-up decisions based on the aggregated weights. Also, the AWGMM has a different aggregation than GMM, especially with respect to the weights of \textit{reliability}, \textit{responsiveness}, and \textit{empathy}. In particular, the weights of \textit{empathy} and \textit{responsiveness} are much less in the AWGMM, while the weight of \textit{reliability} is more significant. To inspect this difference, we looked into the participants' priorities and realized that 7 out of 140 participants had been assigned an infinitesimal weight. By looking into the priorities of these participants, we realize that most of them have either assigned significantly higher weights to \textit{empathy} (mainly more than $0.50$) or \textit{tangibles} and instead a much lower weight to \textit{reliability}. Hence, since these participants are deemed deviants and assigned a lower weight, they have a lesser impact on the final aggregated priorities. As a result, the AWGMM aggregated priorities have a higher weight for \textit{reliability} and lower weights for \textit{tangibles} and \textit{empathy} in comparison to the GMM.

\subsection{Credal ranking}
We now compute the credal ranking of criteria based on the Bayesian Wilcoxon Signed-rank test. We specifically use the Bayesian Wilcoxon Signed-rank test because it entails fewer assumptions on the input data and is typically used in MCDM. To better summarize the credal ranking of criteria, we visualize it using a weighted, directed graph. The nodes in the graph are the criteria, and each directed arc shows that the criterion in origin is much more important than that at the other end by a confidence level specified by the weight of the associated arc. Each arc in the graph visualizes a credal ordering of two criteria, and the whole graph visualizes the credal ranking of all criteria. 

\begin{figure*}[t]
    \centering
    \includegraphics[width=.9\textwidth]{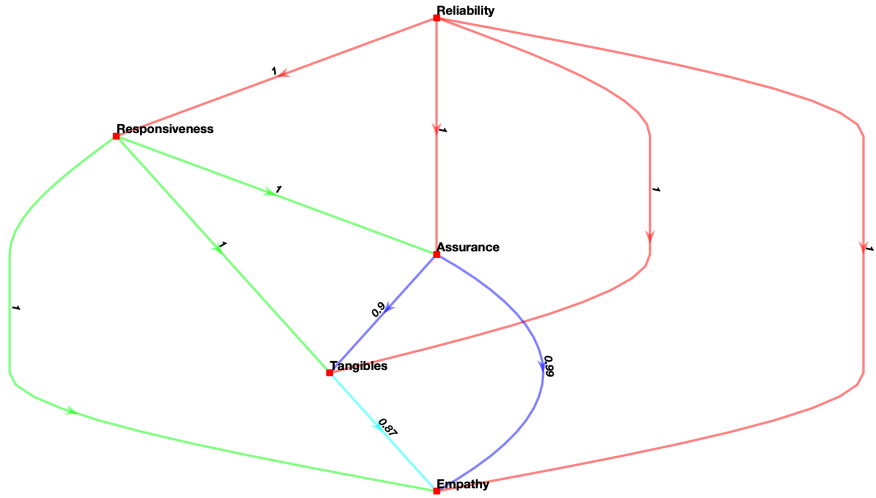}
    \caption{The credal ranking of criteria for assessing the service quality of airline baggage handling system.}
    \label{fig:baggage credal ranking}
\end{figure*}

Figure \ref{fig:baggage credal ranking} shows the credal ranking of five criteria for the baggage handling systems. According to this graph, \textit{reliability} is by far the most important criterion, followed by \textit{responsiveness}. \textit{Assurance} is the third important criterion, and it is more important than \textit{tangible} and \textit{empathy} with confidence levels of 0.90 and 0.99, respectively. Also, \textit{tangible} is the fourth criterion and is more important than \textit{empathy} with a confidence of 0.87. 

\subsection{Clustering the participants}
We now show how clustering can group the participants into multiple mutually exclusive groups. For doing so, we apply\textit{ K}-means for compositional data discussed in Section \ref{sec:distance}, and compare with conventional \textit{K}-means. Also, the number of clusters is set to three \citep{servqual}. We use the mean absolute deviation and its compositional version for clustering to highlight the advantages of the compositional distance metrics for grouping the DMs based on their priorities.

We first compare the outcome of clustering methods based on their centroids. Table \ref{tab:clustering} shows the centroids of clusters identified by the clustering methods. An essential difference between the methods is the difference between the sum of centroids: The centroids of\textit{ K}-means do not sum up to one, while those of the compositional \textit{K}-means will add up to one, satisfying the unit-sum constraint required for the priorities of criteria. As a result, the centers of clusters provided by the proposed method are better representatives of the different groups of participants. 

We also compare the clusters of the participants assigned by different clustering methods. The two clustering methods group the participants differently: By repeating the clustering methods multiple times, the two methods assign around 50 participants into different clusters. Thus, without considering the compositional nature, clustering would group the participants differently. While we cannot compare the participants in the clusters, we should note that the clustering based on the conventional distance metric (without considering the compositional nature) is theoretically incorrect, so we should favor the clusters provided by the proposed clustering algorithm that is specially tailored for the compositional data.

\begin{table*}[t]
    \centering
    \caption{The center of three clusters identified by K-means and K-means for compositional data.}
    \label{tab:clustering}
    \begin{tabular}{c|ccccc|c}
    & Tangibles & Reliability & Responsiveness & Assurance & Empathy & sum\\ \hline 
         \multirow{3}{*}{K-means} &  0.1211 & 0.5019 & 0.1402 & 0.1302 & 0.0651 & 0.9584   \\
         &0.1085 & 0.1434 & 0.4957 & 0.1185 & 0.1142 & 0.9803 \\
         &0.1092 & 0.1545 & 0.1549 & 0.1806 & 0.1765 & 0.7757\\ \hline 
         
         \multirow{3}{*}{Compositional K-means} & 0.1524 & 0.1524 & 0.4653 & 0.0773 & 0.1524 & 1 \\
         & 0.1375 & 0.5325 & 0.1375 & 0.1375 & 0.0550 & 1\\
         & 0.0726 & 0.3224 & 0.1921 & 0.2005 & 0.2123 & 1 \\ \hline 
         \end{tabular}
    
\end{table*}

\section{Conclusion and Discussion}\label{sec:conclusion}
In this paper, we studied three different errors in processing the priorities of multiple decision-makers (DMs) in group decision-making problems, and correct ways for processing the priorities were introduced. The first error discussed in this article was the aggregation of priorities and showed that the compositional analysis for aggregation would result in the normalized geometric mean of priorities. An essential by-product is that the use of the arithmetic mean of priorities should be avoided. 

We also discussed the error regarding the computation of the standard deviation of weights and proposed using Bayesian statistics to provide a probabilistic ranking of criteria, called credal ranking. The credal ranking gives the extent to which a group of DMs prefers one criterion over another, computed based on the Bayesian Wilcoxon Signed-rank test. We finally explained a proper distance metric for gauging the distance between two priorities, according to which we modified\textit{ K}-means clustering for grouping the DMs based on their priorities. 

The findings of this study also have implications for other statistical tools when we use them for priorities (e.g., analysis of variance (ANOVA)). Generally speaking, wherever we need to do computations with priorities, we need to consider their compositional nature. 

In the future, more errors in MCDM should be studied from a compositional data perspective. A crucial case is when the performance matrix of alternatives is also created by some MCDM methods, each column resulting in a composition. Then, the compositions from different criteria should be merged into global priorities by considering the importance of criteria shown by another composition. Considering the compositional nature of the data might allow us to extend those methods to be normalization-agnostic and obviate the rank reversal phenomenon.        

\bibliographystyle{elsarticle-harv}
\bibliography{main}
\end{document}